\documentclass[twocolumn]{revtex4}
\usepackage{graphicx}
\usepackage{amsbsy}

\begin{document}

\title{A general T-matrix approach applied to two-body and three-body problems\\
 in cold atomic gases}
\author{Xiaoling Cui}
\affiliation{Institute for Advanced Study, Tsinghua University,
Beijing, 100084, China}
\date{{\small \today}}
\begin{abstract}

We propose a systematic T-matrix approach to solve few-body problems
with s-wave contact interactions in ultracold atomic gases. The
problem is generally reduced to a matrix equation expanded by a set
of orthogonal molecular states, describing external center-of-mass
motions of pairs of interacting particles; while each matrix element
is guaranteed to be finite by a proper renormalization for internal
relative motions. This approach is able to incorporate various
scattering problems and the calculations of related physical
quantities in a single framework, and also provides a physically
transparent way to understand the mechanism of resonance scattering.
For applications, we study two-body effective scattering in 2D-3D
mixed dimensions, where the resonance position and width are
determined with high precision from only a few number of matrix
elements. We also study three fermions in a (rotating) harmonic
trap, where exotic scattering properties in terms of mass ratios and
angular momenta are uniquely identified in the framework of
T-matrix.

\end{abstract}

\maketitle

\section{Introduction}

Interacting ultracold atoms have gained a lot of research interests
for their interaction strength and dimensionality are highly
controllable by making use of Feshbach resonance and external
confinements\cite{Bloch,Giorgini}. In such dilute atomic gases, the
interaction between atoms can be well approximated as contact
potential which is characterized by the s-wave scattering
length\cite{Bloch,Giorgini}. In this context, the few-body problems
play very important roles in studying many-body properties. For
instance, solutions of these problems determine effective
interactions between atom-atom, atom-dimer and
dimer-dimer\cite{Petrov1, Petrov2}, which are fundamental elements
to formulate the many-body effective Hamiltonian; moreover, the
consideration of two-body short-range physics leads to a series of
exact universal relations for a many-body system, as first proposed
by Tan\cite{Tan1, Tan2, Tan3} and recently verified in
experiment\cite{Jin10}.

Previous studies of few-body systems have revealed many nontrivial
effects. One famous example is the Efimov effect for three
atoms\cite{Efimov,Braaten06}, depending closely on the short-range
interacting parameter apart from single s-wave scattering length.
Another typical effect is the two-body induced resonance scattering
and induced bound state due to external confinements\cite{Busch98,
Olshanii1, Olshanii2, Moritz05, Petrov00, Petrov01, Orso05,
Buechler10, Cui10, Kestner07, Peano, Castin,Nishida_mix, Lamporesi}.
Among most of previous studies, the problems were solved in the
framework of two-channel models\cite{Kokkelmans02} or by using
pseudopotentials\cite{LHY1,LHY2}. 
In this article, we present using T-matrix approach to solve
few-body problems with s-wave contact interaction in the field of
ultracold atoms. 
Compared with other methods, T-matrix is able to systematically
provide exact solutions for few-body problems, and more importantly,
is able to work in a much efficient and physically transparent way.

In this article, we shall first formulate T-matrix method and
introduce its essential concept, i.e., the {\em renormalization}
idea to integrate out all high-energy(or short-range) contributions
for relative motions. Then a systematic treatment is presented to a
general N-body system with contact interactions and with possibly
trapping potentials. The key point of this method is to make use of
the interaction property and introduce a set of orthogonal molecular
states, which describe the {\em external} center-of-mass(CM) motions
of pairs of interacting particles; then the problem is generally
reduced to a matrix equation, and a proper renormalization scheme
for the {\em internal} relative motions ensures finite value of each
matrix element. Using this method, we obtain the bound state
solution, scattering matrix element and reduced coupling constant in
the low-dimensional subspace. Moreover, the treatment has a lot more
physical meanings and allows us to make analytical predictions to
various induced resonances under confinement potentials. In all,
T-matrix approach is able to unify many studies of different issues
in the single framework.

To show the efficiency of this approach, we apply it to study
two-body effective scattering in 2D-3D mixed dimensions, where
T-matrix not only provides a transparent way to understand the
mechanism of multiple resonances, but also gives explicit
expressions for the resonance position and the width. Particularly,
its efficiency lies in that each resonance can be determined
accurately by only calculating a few number of matrix elements.
Moreover, we apply this method to study three two-component fermions
in a (rotating) harmonic trap, and show its unique advantage in
identifying scattering properties in different angular momentum
channels. The ground state is obtained for a rotating and trapped
system, which gives important hints for quantum Hall physics in the
fermionic atom-dimer system.

The rest of this paper is organized as follows. In section II, we
introduce the renormalization concept and present T-matrix formulism
to solve a general few-body problem. Its relations to other
approaches, advantages and limitations are also discussed. Section
III is the application of T-matrix approach to two-body problems,
where specifically we study the scattering resonances in 2D-3D mixed
dimensions. Section IV is the application to three fermions in a
(rotating) harmonic trap, where the energy spectrum and scattering
property are studied for different mass ratios and angular momenta
of three fermions. We summarize the paper in the last section.

\section{T-matrix approach}

In this section we give a systematic prescription of T-matrix
approach to solve few-body problems. The resulted matrix equation is
given by Eq.\ref{f_matrix} in Section IIA, from which we extract
three important physical quantities as given by
Eqs.(\ref{Eb},\ref{T-matrix},\ref{geff}) in Section IIB. We also
discuss its relation to other widely used methods in Section IIC,
and demonstrate its unique advantages and limitations in Section
IID.

\subsection{Basic concept and general formulism}

We start from the Lippmann-Schwinger equation based on standard
scattering theory,
\begin{eqnarray}
|\psi\rangle&=&|\psi_0\rangle+\hat{G}_0(E)\hat{T}|\psi_0\rangle.\label{psi}
\end{eqnarray}
Here $|\psi_0\rangle$ is the eigen-state for non-interacting system,
with Hamiltonian $\hat{H}_0=\hat{H}_{{\rm kin}}+\hat{V}_T$ composed
by kinetic term and external trapping potential; $|\psi\rangle$ is
the scattered state in the presence of interaction potential
$\hat{U}$;
\begin{equation}
\hat{G}_0(E)=\frac{1}{E-\hat{H}_0+i\delta} \label{G}
\end{equation}
is the Green function; the scattering matrix can be expanded as
series $T=U+UG_0U+UG_0UG_0U+...$ which leads to
\begin{eqnarray}
T&=&(1-UG_0)^{-1}U . \label{T}
\end{eqnarray}
To this end, $G_0$, $T$ and $U$ are all matrixes expanded by certain
complete set of states. If we use $\{|\psi_0\rangle\}$ to expand
T-matrix, then each T-matrix element in Eq.\ref{T} directly relates
to the scattering amplitude in the scattered wavefunction and
therefore represents the effective interaction in the low-energy
space. To obtain the effective interaction, a physically insightful
way is to employ the concept of {\em renormalization}. For the
fundamental two-body s-wave scattering with contact interaction,
\begin{equation}
\hat{U}(r)=U_0\delta^3(\mathbf{r}), \label{U0}
\end{equation}
one can
resort to a simple momentum-shell renormalization
scheme\cite{Cui10}. The spirit here is to take all intermediate
scattering from low-$\mathbf{k}$(momentum) space to the shell in
high-$\mathbf{k}$ space as virtual processes, which in turn modify
the effective interaction strength($U$) in the low-$\mathbf{k}$
space by a small $\delta U$; perturbatively $\delta U$ in terms of
the shell momentum($\delta\Lambda$) follows
\begin{equation}
\delta U=\sum_{\Lambda-\delta\Lambda<|{\bf k}|< \Lambda
}-\frac{U^2}{V}\frac{1}{\epsilon_{\bf k}},
\end{equation}
here $V$ is the volume, and $\epsilon_{\bf k}={\bf k}^2/(2\mu)$ is
the energy for relative motion of two particles with reduced mass
$\mu$.
The resultant RG flow equation reads\cite{Kaplan}
\begin{equation}
\frac{1}{U^2}\frac{\delta U}{\delta
\Lambda}=\frac{1}{V}\frac{\delta}{\delta \Lambda}(\sum_{|{\bf k}|<
\Lambda }\frac{1}{\epsilon_{\bf k}})
\end{equation}
relates the bare interaction $U_0$ to the zero-energy effective
one($T_0=2\pi a_s/\mu$) via,
\begin{equation}
\frac{\mu}{2\pi a_s}=\frac{1}{U_0}+\frac{1}{V}\sum_{\mathbf{k}}
\frac{1}{\epsilon_{\mathbf{k}}}, \label{freeRG}
\end{equation}
with $a_s$
the s-wave scattering length. 
For a general N($>2$)-body problem, the idea of renormalization,
though not as explicitly shown as above, always serves as the
underlying principle through the whole scheme(see below).

Now we proceed with the general T-matrix approach. Suppose a
$Q-$species system, and the $i$-th ($i=1...Q$) species has $N_i$
identical particles residing at
$\mathbf{x}^i_{1},...\mathbf{x}^i_{N_i}$; $U_{i}$ and
$U_{ij}(i<j<Q)$ are respectively the bare interaction strengths
between particles within the $i-$th species and between different
species ($i$ and $j$); $U_{i}$ ($U_{ij}$) is related to the
corresponding scattering length $a_i$ ($a_{ij}$) via
Eq.\ref{freeRG}. Taking advantage of the zero-range property of the
interaction
\begin{equation}
\hat{U}=\sum_{i=1}^Q\sum_{m<n}^{N_i}U_i\delta^3(\mathbf{x}^i_m-\mathbf{x}^i_n)+\sum_{i<j}^{Q}\sum_{m=1}^{N_i}\sum_{n=1}^{N_j}U_{ij}\delta^3(\mathbf{x}^i_m-\mathbf{x}^j_n),
\label{U}
\end{equation}
and thus the same property of T-matrix given by Eq.\ref{T}, we
expand $U$ and $T$ by a set of molecular states
$\{|\mathbf{x}^i_m-\mathbf{x}^j_n=\mathbf{0},\lambda\rangle\}$. This
state is defined such that one pair of interacting
particles($\mathbf{x}^i_m, \ \mathbf{x}^j_n$) locate at the same
site, and $\lambda$ is the energy-level index for the remanent
degrees of freedom. More detailed description of the molecular state
is given in Appendix \ref{indi mole}. Further, for identical
bosons/fermions the molecular state should further be
symmetrized/antisymmetrized by superpositions of above individual
ones. Explicitly we have
\begin{equation}
\hat{T}|\psi_0\rangle=\sum_{I\lambda}f^I_{\lambda}|r_{I}=0,\lambda\rangle,
\label{f}
\end{equation}
here each state,
$|r_I=0,\lambda\rangle$($r_I=|\mathbf{x}^i_m-\mathbf{x}^j_n|$) with
$I\leq\max(I)= \frac{Q(Q+1)}{2}$, corresponds to one interaction
term in Eq.\ref{U}. The coefficients $\{f^I_{\lambda}\}$ satisfy
\begin{eqnarray}
\sum_{I'\lambda'}f^{I'}_{\lambda'}&\big[&\frac{\mu_I }{2\pi
a_I}\delta_{II'}\delta_{\lambda\lambda'}-C^{II'}_{\lambda\lambda'}\big]
=\langle r_I=0,\lambda|\psi_0\rangle \label{f_matrix}
\end{eqnarray}
with
\begin{equation}
C^{II'}_{\lambda\lambda'}=\frac{1}{V}\sum_{\mathbf{k}}
\frac{1}{\epsilon_{\mathbf{k}}}\delta_{II'}\delta_{\lambda\lambda'}
+\langle r_I=0,\lambda|\hat{G}_0|r_{I'}=0,\lambda' \rangle.
\label{f_matrix2}
\end{equation}
Detailed derivation of Eq.\ref{f_matrix} is given in Appendix
\ref{C_derive}.

For the two-particle scattering in free space, the energy level
$\{\lambda\}$ is characterized by the CM momentum, which is
conserved by the interaction and thus irrelevant to the scattering
problem for relative motions. The molecular state is then as simple
as $|r=0\rangle$, with $r$ the distance between two particles.
Eq.\ref{f} is then reduced to $\hat{T}|\psi_0\rangle=f|r=0\rangle$,
and $f$ is given by Eq.\ref{f_matrix} as $f^{-1}\propto
a_s^{-1}+ik$, reproducing the well-known relation between the
scattering amplitude and the s-wave scattering length ($a_s$). The
applications of this approach to other few-body systems will be
introduced in Section III and IV.

\subsection{Calculation of physical observables}

With the information of molecular states, 
all physical quantities can be deduced straightforwardly. We shall
enumerate below three quantities that are detectable or observable
in experiments.

{\em (I)Bound state solution.} In this case $|\psi_0\rangle$ is
absent, Eq.\ref{f_matrix} is given by the pole of T-matrix(see
Eq.\ref{T}), i.e.,
\begin{equation}
{\rm Det}(1-UG_0(E_b))=0, \label{Eb}
\end{equation}
where $E_b$ is the binding energy, and the eigen-vector
$\{f^{I}_{\lambda}\}$ gives the bound state as {\rm
\begin{equation}
|\psi_b\rangle=\sum_{I\lambda}f^{I}_{\lambda}\hat{G}_0(E_b)|r_I=0,\lambda\rangle.\label{psi_b}
\end{equation}
For two-particle scattering(with scattering length $a_s$) in free
space, Eqs.(\ref{Eb}, \ref{psi_b}) give $E_b=-1/(2\mu a_s^2)$ and
$|\psi_b\rangle\propto \sum_{\mathbf{k}}
\frac{1}{E_b-\epsilon_{\mathbf{k}}}|\mathbf{k}\rangle$.

{\em (II)T-matrix element.} Generally, T-matrix element between
$|\psi_0\rangle$ and itself characterizes the scattering property of
low-energy particles. With Eqs.(\ref{f},\ref{f_matrix}), we obtain
\begin{equation}
\langle\psi_0|\hat{T}|\psi_0\rangle=\sum_{I\lambda}f^I_{\lambda}\langle
\psi_0|r_{I}=0,\lambda\rangle,\label{T-matrix}
\end{equation}
which can also be obtained from Eq.\ref{T} as
$\langle\psi_0|(1-UG_0)^{-1}U|\psi_0\rangle$. For two-particle
scattering in free space, $f^I_{\lambda}\equiv f$ is proportional to
the scattering amplitude.

{\em (III)reduced interaction.} If trapping potentials confine atoms
in a lower dimension, there are two distinct scattering channels for
the low-energy state, namely the open(P) or closed(Q) channel,
depending on whether its wavefunction propagates or decays at large
interparticle distance in the lower dimension. The effective
interaction strength for low-energy particles in the open channel is
modified from the original bare one by virtual scatterings to the
closed channels. We assume $g_{\rm eff}$ as the modified interaction
strength in open channel, which, for instance, has been defined in
the reduced 1D Hamiltonian\cite{Olshanii1, Olshanii2, Peano, note}
under tight transverse harmonic traps. Following the same procedure
in obtaining T-matrix element in (II), we have
\begin{equation}
g_{\rm
eff}=U_{PP}+U_{PQ}G_0^{Q}U_{QP}+U_{PQ}G_0^{Q}U_{QQ}G_0^{Q}U_{QP}+...,
\label{g_eff}
\end{equation}
where $U_{ts}(t,s=\{P,Q\})$ is the bare coupling strength between
two particular channels;
\begin{equation}
G_0^Q=\frac{1}{E-H_0^Q+i\delta}\label{GQ}
\end{equation}
is the Green function for Hamiltonian $H_0^Q$ that only acts on
closed-channel states. Again due to the property of
$\hat{U}$(Eq.\ref{U}), we insert into Eq.\ref{g_eff} a set of
molecular states $\sum_{I\lambda}|r_I=0,\lambda\rangle\langle
r_I=0,\lambda|$\cite{insert}. Assuming two column vectors
\begin{equation}
\xi=\{\langle r_I=0,\lambda|\psi_0\rangle\},\ \ \zeta=\{\langle
r_I=0,\lambda|\hat{U}\psi_0\rangle\},
\end{equation}
we then obtain
\begin{eqnarray}
g_{\rm
eff}&=&{\xi}^T\zeta+{\xi}^T(UG_0^Q)\zeta+{\xi}^T(UG_0^Q)^2\zeta+...\nonumber\\
&=&\xi^T(1-UG_0^Q)^{-1}\zeta\nonumber\\
&=&\langle\psi_0|(1-UG_0^Q)^{-1}U|\psi_0\rangle,\label{geff}
\end{eqnarray}
here $UG_0^Q$ is a matrix expanded by molecular states. Note that
$g_{\rm eff}$ is different from
$\langle\psi_0|\hat{T}|\psi_0\rangle$ in (II) only by the Green
function therein. Obviously $g_{\rm eff}$ is the renormalized
coupling strength in the open channel by all virtual scattering to
states in closed channels.

\subsection{Relations with other methods}

In this section, we analyze the intrinsic relation between T-matrix
and other widely used methods, such as those in the framework of
two-channel models\cite{Buechler10, Kestner07} and
pseudopotentials\cite{Busch98, Olshanii1, Olshanii2, Orso05, Duan07,
Liu10}.

In two-channel models, the closed-channel molecules are explicitly
included in the Hamiltonian; these molecules couple to atoms in
open-channel and thus mediate interactions between the atoms. In the
present T-matrix method, the molecular states(defined in Eq.\ref{f})
can be considered as the analog of closed-channel molecules in
two-channel models. The similarities lie in that they are both
constructed in a way that follows the zero-range property of
interaction, and describe the CM motion of two interacting
particles.

In pseudopotentials,
the problem is solved by applying Bethe-Peierls boundary condition
to the wavefunction at short inter-particle distance, i.e.,
\begin{equation}
\lim_{r_{I}=|\mathbf{x}_1-\mathbf{x}_2|\rightarrow0}\psi(\mathbf{x}_1,\mathbf{x}_2,\mathbf{x}_3...)
=(\frac{1}{r_{I}}-\frac{1}{a_{I}})f(\mathbf{x}_1=\mathbf{x}_2,\mathbf{x}_3...),\label{asymp}
\end{equation}
here the index $I$ denotes the pair $\{\mathbf{x}_1, \
\mathbf{x}_2\}$ and $a_{I}$ is the scattering length between them.
On the other hand, we notice that above asymptotic behavior can be
automatically satisfied by the present scheme of T-matrix method.
To show this, we examine Eq.\ref{psi} at short inter-particle
distance by projecting it to certain molecular state,
\begin{eqnarray}
\lim_{r_I\rightarrow 0} \langle r_I,\lambda |\psi \rangle &=&\langle
r_I=0,\lambda|\psi_0\rangle+\nonumber\\
&&\sum_{I'\lambda'}f^{I'}_{\lambda'}\langle
r_I=0,\lambda|\hat{G}_0|r_{I'}=0,\lambda\rangle \nonumber\\
&=&f^{I}_{\lambda}(\frac{\mu_I }{2\pi
a_I}-\frac{1}{V}\sum_{\mathbf{k}}\frac{1}{\epsilon_{\mathbf{k}}})\nonumber\\
&=&\lim_{r_I\rightarrow 0}f^{I}_{\lambda}\frac{\mu_I
}{2\pi}(\frac{1}{a_I}-\frac{1}{r_I}). \label{univ}
\end{eqnarray}
To derive Eq.\ref{univ} we have used Eq.\ref{f_matrix} and the
Fourier transformation of zero-energy Green function in free space.
Remarkably, Eq.\ref{univ} shows that the asymptotic form is given by
the inverse of bare potential $1/U_I$, which is universal regardless
of any trapping potential. In this sense, the Bethe-Peierls boundary
condition (Eq.\ref{asymp}) in the framework of pseudopotentials is
equivalent to renormalization equation (Eq.\ref{freeRG}) in T-matrix
method.

\subsection{Advantages and Limitations}

In principle, the T-matrix fomulism presented in Section IIA is
appliable to a general few-body problem with zero-range
interactions. It gives a unified treatment to different scattering
issues, as shown in Section IIB. Besides, T-matrix approach has the
following advantages.

First, we compare this scheme, using molecular states to expand
T-matrix (Eq.\ref{f_matrix}, with that using original $N-$particle
states. The latter requires a matrix dimension as large as
$\Gamma^{N}$ ($\Gamma$ is the cutoff index for single particle
energy levels), while the former could reduce it to at most
$\frac{Q(Q+1)}{2}\Gamma^{N-1}$ for a general case and further to
$\frac{Q(Q+1)}{2}\Gamma^{N-2}$ for the special case when CM motion
can be separated out. 

Second, this scheme is physically insightful in that it reveals
general couplings between the CM and relative motions. Specifically,
CM serves as {\em external} indices of the matrix, while relative
motions contribute to each element by the renormalization of {\em
internal} degree of freedom. As we shall see in Section III, this
picture is essentially important to understand the mechanism of
resonance scattering in the two-body system.

More advantages related to the realistic calculations will be shown
in Section IIIB and IV, when applying this method to specific
two-body and three-body problems.

However, the present T-matrix method still has limitations under
certain circumstances. When examining
Eqs.(\ref{f_matrix},\ref{f_matrix2}), one can see that the
convergence of the solution from the matrix equation generally
requires two conditions. First, each matrix element is finite;
secondly, the solution is independent of the matrix size. Any
violation of above two conditions implies that the zero-range model
is insufficient to characterize the interacting system, such as when
Efimov physics emerges in the three-body sector\cite{Efimov,
Braaten06}. T-matrix is able to identify the violation of the first
condition (see Section IV). However, it would be quite involved for
it to identify the second condition. Alternatively, when Efimov
physics appear, one can solve the problem by resorting to
hyperspherical coordinate method in unitary
limit\cite{Werner061,Werner062}, or by employing a three-body force
to eliminate the cutoff dependence\cite{Bedaque}. The extension of
the present T-matrix approach to identify such nontrivial few-body
effects is out of the scope of this paper.

\section{Application to two-body problem}

In this section we consider the two-body system. First we present
the formulism of physical quantities introduced in Section IIB, for
both cases when the CM motion and relative motion can or cannot be
decoupled. Finally we apply this method to address the effective
scattering of two particles in 2D-3D mixed dimensions. Particularly
we emphasize the physical insight given by T-matrix to understand
the resonance mechanism, as well as its efficiency in realistic
calculation of scattering parameters.

\subsection{Formulism}

\subsubsection{$\mathbf{r}$ and $\mathbf{R}$ decoupled system}

For trapping potential
$V_T(\mathbf{x}_1,\mathbf{x}_2)=V_T(\mathbf{R})+V_T(\mathbf{r})$,
the molecular state is simply $|r=0,\lambda_0\rangle$, with
$\lambda_0$ characterizing the CM motion and staying unaffected by
the interaction. The bound state solution $E_b$ is determined from a
single equation as
\begin{equation}
\frac{\mu}{2\pi a_s}=C(E_b),\label{E_b}
\end{equation}
with
\begin{eqnarray}
C(E)&=&\frac{1}{V}\sum_{\mathbf{k}}
\frac{1}{\epsilon_{\mathbf{k}}}+\sum_l
\frac{|\phi_l(0)|^2}{E-E_l+i\delta},\label{CE}
\end{eqnarray}
here $\{\phi_n(\mathbf{r})\}$ is a complete set of eigen-states only
for the relative motion. A convergent solution of $E_b$ requires
that the ultraviolet divergence in each term of Eq.\ref{CE} be
exactly cancelled with each other. This is actually satisfied by a
regular potential $V_T(\mathbf{x})$ without singularity at any
position $\mathbf{x}$\cite{convergence}.

The T-matrix element, $T_{mn}\equiv\langle m|\hat{T}(E)|n\rangle$,
which represents the scattering amplitude from initial state
$|n\rangle$ to final state $|m\rangle$, is given by
\begin{equation}
\frac{\phi^*_m(0)\phi_n(0)}{T_{mn}}=\frac{\mu}{2\pi a_s}-C(E).
\label{TE}
\end{equation}
Compared with Eq.\ref{E_b}, this shows that a bound state emerges
when {\em all} elements $T_{mn}$ simultaneously diverge.

Similarly for reduced interaction in the lower dimension, we obtain
\begin{equation}
\frac{|\psi_0(0)|^2}{g_{\rm eff}}=\frac{\mu}{2\pi
a_s}-C^Q(E),\label{gE}
\end{equation}
where $C^Q$ follows the form of Eq.\ref{CE} with the second
summation over all closed channel states. In this sense the
confinement induced resonance(CIR), referring to $g_{\rm
eff}\rightarrow\infty$, occurs at
\begin{equation}
\frac{\mu}{2\pi a_s}=C^Q(E).
\end{equation}
This in turn determines a closed channel bound state with the same
energy as $|\psi_0\rangle$. Compared to existing exploration of CIR
mechanism in quasi-1D system\cite{Olshanii1, Olshanii2}, T-matrix
shows in a general way how the divergence of reduced interaction in
the open channel is associated with the emergence of a
closed-channel bound state at the same energy level.

Finally, Eqs.(\ref{CE},\ref{TE},\ref{gE}) indicate the way how the
trapping potentials modify the low-energy scattering theory. The
modification is in fact through the intermediate virtual scattering
processes, i.e., by redistributing the energy levels and changing
coupling strengths between these states. Above formula can be
applied to ordinary harmonic confinements studied
before\cite{Busch98, Olshanii1, Olshanii2, Petrov00, Petrov01,
Kestner07}.

\subsubsection{$\mathbf{r}$ and $\mathbf{R}$ coupled system}

For a general trapping potential, the two-body non-interacting
Hamiltonian can be divided to three pieces, describing the relative
motion $H_{rel}(\mathbf{r})$, CM motion $H_{cm}(\mathbf{R})$, and
couplings in-between $H_{cp}(\mathbf{r},\mathbf{R})$. The molecular
state is then introduced as $|r=0,\lambda\rangle$, and
$\Phi_{\lambda}(\mathbf{R})\equiv\langle \mathbf{R}|\lambda\rangle$
is the eigen-state of
\begin{equation}
H_{cm}(\mathbf{R})=-\frac{\nabla^2_{\mathbf{R}}}{2M}+V_{T,1}(\mathbf{R})+V_{T,2}(\mathbf{R}).
\end{equation}
Combining with Eq.\ref{f_matrix}, one can obtain all solutions
corresponding to (I,II,III) in Section IIB.

In this case, the trapping potential can induce multiple two-body
scattering resonances as revealed previously in several
settings\cite{Peano,Nishida_mix,Lamporesi} by numerical calculations
in coordinate space. Next we show that these resonances can be
analytically figured out in the framework of T-matrix method. We
classify the situations by whether the effective scattering is in 3D
space or in the reduced lower dimension.

When external trapping potentials are applied but the low-energy
scattering wavefunction, $\psi(\mathbf{x}_1,\mathbf{x}_2)$, can
still behave in a propagating way at large 3D inter-particle
separations $|\mathbf{x}_1-\mathbf{x}_2|\rightarrow\infty$, then its
asymptotic form can be written as
\begin{equation}
\psi(\mathbf{x}_1,\mathbf{x}_2)\sim1-\frac{a_{\rm
eff}}{d(\mathbf{x}_1,\mathbf{x}_2)}.
\end{equation}
Here $d$ is the modified interparticle distance according to the
confinement(see also Ref.\cite{Nishida_mix} and discussions in
Appendix \ref{a_eff}); $a_{\rm eff}$ is the effective scattering
length, and can be directly related to $T-$matrix element for
zero-energy scattering state\cite{aeff-T_proof}.

According to T-matrix method, assuming Eq.\ref{f_matrix2} yields
\begin{equation}
C_{\lambda\lambda'}\mathcal{F}_{\lambda'\nu}=c_{\nu}\mathcal{F}_{\lambda\nu},
\end{equation}
with $c_{\nu}$ ($\nu=1,2...)$ the eigenvalue and $\mathcal{F}_{\nu}$
the corresponding eigenvector, then we have
\begin{equation}
\langle\psi_0|T|\psi_0\rangle=\sum_{\nu}\frac{\langle\psi_0|\mathcal{F}_{\nu}\rangle\langle
\mathcal{F}_{\nu}|\psi_0\rangle}{\frac{\mu }{2\pi
a_s}-c_{\nu}}.\label{T-aeff}
\end{equation}
This equation explicitly predicts an infinite number of resonances
($a_{\rm eff}\rightarrow\infty$) when each discretized $c_{\nu}$
individually match with $\frac{\mu }{2\pi a_s}$ by tuning $a_s$ in
realistic experiments. The resonance position and width can be
conveniently extracted from the exact diagonalization of C-matrix.

To explore the mechanism of these resonances, first we only focus on
the diagonal elements of C-matrix. Within each molecular channel,
all relative motion levels are coupled together by the attractive
interaction and this potentially leads to a bound state. This bound
state(relative motion) combined with the molecular channel(CM
motion) tend to produce the zero total energy, and thus give rise to
the divergent T-matrix or $a_{\rm eff}\rightarrow\infty$. By tuning
the interaction or $a_s$, the zero-energy state will emerge in order
from each molecular channel and cause the resonance of $a_{\rm
eff}$. The width of each resonance is determined by the coupling
between the zero-energy scattering state and each molecular state,
which becomes narrower for higher levels of molecular states.

However, above understanding of multiple resonances is not rigorous,
because different molecular channels could also couple with each
other by the combination of $H_{rel}(\mathbf{r})$ and
$H_{cp}(\mathbf{r},\mathbf{R})$. This additional coupling, as shown
by off-diagonal elements of C-matrix, would give a correction to the
ideally predicted resonance position. In the following section, we
shall address these issues by studying a specific system with
multiple resonances, i.e., two atoms scattering in 2D-3D mixed
dimension.

When external trapping potentials are applied such that at low
energies, $\psi(\mathbf{x}_1,\mathbf{x}_2)$ only propagates as two
particles are far apart in a lower dimension(open channel), then the
same analysis can be applied to the effective interaction $g_{\rm
eff}$ in this channel. Now C-matrix is defined by the closed-channel
Green function $G_0^Q$(Eq.\ref{GQ}), which equally results in the
matrix equation
\begin{equation}
C^Q_{\lambda\lambda'}\mathcal{F}^Q_{\gamma\lambda'}=c^Q_{\gamma}\mathcal{F}^Q_{\gamma\lambda}.
\end{equation}
Combined with Eq.\ref{geff}, it gives
\begin{equation}
g_{\rm
eff}=\sum_{\gamma}\frac{\langle\psi_0|\mathcal{F}^Q_{\gamma}\rangle\langle
\mathcal{F}^Q_{\gamma}|\psi_0\rangle}{\frac{\mu }{2\pi
a_s}-c^Q_{\gamma}}.
\end{equation}
Therefore $g_{\rm eff}$ will go through a resonance as long as one
$c^Q_{\beta}$ is matched with $\frac{\mu }{2\pi a_s}$ by tuning
$a_s$. This corresponds to the energy of dressed bound state in each
closed molecular channel moves downwards and touches the threshold
energy of open channel. The resonance width would be narrower for
higher molecular channels due to the smaller overlap with the
low-energy scattering state in open channel.

\subsection{Results of scattering in 2D-3D mixed dimensions}

We consider one atom ($^{41}$K or $^{40}$K, labeled by A) is axially
trapped by a tight harmonic potential with frequency $\omega_A$,
while the other atom ($^{87}$Rb or $^6$Li, labeled by B) is free in
3D space. The Hamiltonian reads
\begin{equation}
H(\mathbf{r}_A,\mathbf{r}_B)=-\frac{\nabla^2_{\mathbf{r}_A}}{2m_A}+\frac{1}{2}m_A
\omega_A^2z_A^2-\frac{\nabla^2_{\mathbf{r}_B}}{2m_B}+U_0\delta^3(\mathbf{r}_A-\mathbf{r}_B).
\end{equation}
As shown in Appendix \ref{a_eff}, the effective scattering length
$a_{\rm eff}$ for zero-energy scattering is defined by the two-body
wavefunction when
$d_{AB}=\sqrt{\frac{\mu}{m_B}\boldsymbol\rho_{AB}^2+z_B^2}\rightarrow\infty$,
\begin{eqnarray}
\psi(\boldsymbol\rho_{AB},z_A,z_B)\rightarrow
\phi_0(z_A,a_0)(1-\frac{a_{\rm eff}}{d_{AB}}),   \label{psi_2d3d}
\end{eqnarray}
with
\begin{equation}
\frac{2\pi a_{\rm
eff}}{\mu}=V\langle\psi_0|T|\psi_0\rangle.\label{aeff-T}
\end{equation}
Here $\phi_n(z,a_0)$ is the eigen-state of 1D harmonic oscillator
with characteristic length $a_0=\sqrt{1/(m_A\omega_A)}$; $\mu=m_A
u/(1+u)$ is the reduced mass and $u=m_B/m_A$ is mass ratio.

The molecular state in this case is $|r=0,N\rangle$, where
$|N\rangle$ denotes an eigen-state of the following Hamiltonian
\begin{equation}
H_{cm}(Z)=-\frac{1}{2(m_A+m_B)}{\frac{\partial^2}{\partial
Z^2}}+\frac{1}{2}m_A\omega_A^2 Z^2, \label{Hcm}
\end{equation}
which describes the CM motion along (trapped) $z$ direction, with
oscillation frequency $\overline{\omega}=\omega_A/\sqrt{1+u}$ and
characteristic length $\overline{a}=a_0/(1+u)^{1/4}$. According to
Eqs.(\ref{T-aeff},\ref{aeff-T}) we obtain
\begin{equation}
\frac{a_{\rm eff}}{a_0}=\sum_{N=0,2,...}\frac{W_N}{\frac{a_0}{
a_s}-e_{N}}, \label{aeff-as}
\end{equation}
here $e_N$ is the eigen-value of $\tilde{C}-$matrix
(Eq.\ref{C_simp}) determined by
$\tilde{C}_{NM}\tilde{\mathcal{F}}_{MN'}=e_{N'}\tilde{\mathcal{F}}_{NN'}$;
the resonance width is given by
\begin{equation}
W_N=|\sum_{N'}\tilde{\mathcal{F}}_{N'N}f_{N';0,0}|^2,
\end{equation}
with $f(N; n_A, k_z)$ defined by Eq.\ref{f-function}. Due to the
contact interaction and reflection symmetry of trapping potential,
only molecular states with even parity ($N=0,2,4...$) are relevant
in this case. Fig.1(a) and Fig.2(a) shows the first five resonances
of $a_{\rm eff}/a_0$ for $^{41}$K-$^{87}$Rb($m_A<m_B$) and
$^{40}$K-$^{6}$Li($m_A>m_B$) mixtures, when tuning $a_0/a_s$ from
the weak coupling($-\infty$) to strong coupling($+\infty$) side. The
($\frac{N}{2}+1$)-th resonance of $a_{\rm eff}$ is characterized by
the position $(a_0/a_s)_{res}=e_N$ and the width $W_N$.

Amazingly, we find good accordance between $e_N$ and each diagonal
matrix elements $\tilde{C}_{NN}$, as shown by Fig.1(b) and Fig.2(b).
That means the correction caused by off-diagonal couplings between
different molecular channels are actually negligible. There are
mainly two reasons for this. First, the amplitudes of off-diagonal
$C-$matrix elements ($\widetilde{C}_{NN'}$) are much smaller than
diagonal ones, and decrease rapidly as $|N-N'|$ increases. Secondly,
it can be attributed to the destructive interference among couplings
with different molecular channels. To see this, we carry out a
perturbative calculation in terms of off-diagonal couplings between
neighboring molecular states, and get the relative correction to the
$N-$th eigen-value as
\begin{equation}
\Delta_N=\sum_{N'=N\pm
2}\frac{|\tilde{C}_{NN'}|^2}{\tilde{C}_{NN}-\tilde{C}_{N'N'}}/\tilde{C}_{NN}.\label{Delta}
\end{equation}
Note that $N-2$ and $N+2$ terms contribute to $\Delta_N$ with
opposite signs, which results in further suppressed $\Delta_N$.
Insets of Fig.1(b) and Fig.2(b) show $\Delta_N$ for the first eleven
resonances; we see that the most significant effect of off-diagonal
couplings occurs for the first resonance but is still negligibly
small ($\sim1.5\%$ for K-Rb mixture and $\sim0.1\%$ for K-Li
mixture).

The vanishing off-diagonal couplings between different molecular
channels establish the unique advantage of T-matrix scheme, i.e.,
the resonance can be accurately determined by only a few number of
related matrix elements. In the limit of zero couplings, we have
$\tilde{\mathcal{F}}_{N'N}\approx\delta_{N'N}$ and therefore
$e_N=\widetilde{C}_{NN}$, $W_N=|f_{N;0,0}|^2$(see
Eq.\ref{f-function}). 
In this limit and particularly for resonances at $a_s>0$ side, the
resonance position can be determined by matching bound state energy
in each CM molecular channel with the threshold energy of two
particles, i.e.,
\begin{equation}
(N+\frac{1}{2})\overline{\omega}-\frac{1}{2\mu
a_s^2}=\frac{\omega_A}{2},\ \ \ (N=2,4,6...)\label{fit}
\end{equation}
as shown by the dashed lines in Fig.1(b) and Fig.2(b). This
equation, as is the direct outcome of T-matrix method, has been used
previously to determine the resonance positions\cite{Lamporesi}.

\begin{figure}[ht]
\includegraphics[height=5.5cm,width=8cm]{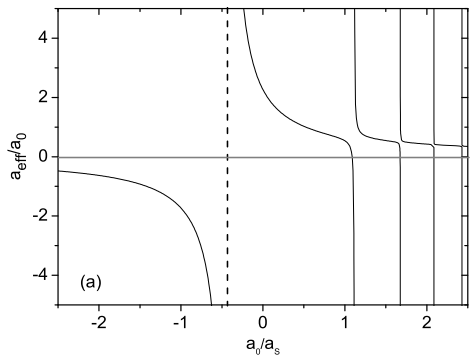}\\
\includegraphics[height=5.5cm,width=7.5cm]{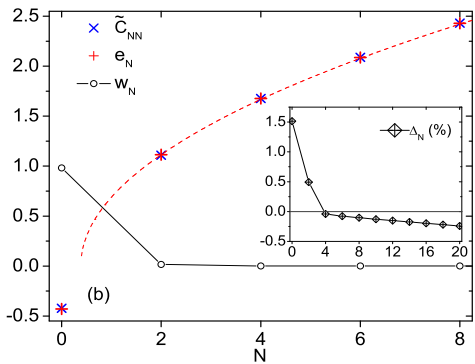}
\caption{(Color online) (a).Effective scattering length $a_{\rm
eff}/a_0$ as functions of $a_0/a_s$ for $^{41}$K(2D)-$^{87}$Rb(3D)
mixture. The dashed line denotes the first resonance at
$a_0/a_s=-0.43$. (b)Diagonal matrix element $\tilde{C}_{NN}$ of
Eq.\ref{C_simp}(denoted by $\times$), the corresponding resonance
position $(a_0/a_s)_{res}$($+$) and resonance width($\circ$). Red
dashed line is the function fit according to Eq.\ref{fit}. Inset
shows the relative correction $\Delta_N$ as defined in
Eq.\ref{Delta}. $N=0,2,4...$ in (b) respectively correspond to the
$(N/2+1)-$th induced resonance in (a) from left to right. }
\label{fig1}
\end{figure}

\begin{figure}[ht]
\includegraphics[height=5.5cm,width=8cm]{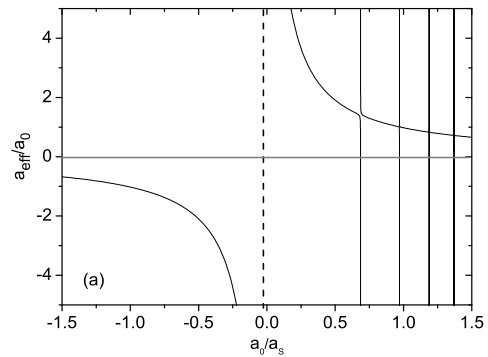}\\
\includegraphics[height=5.5cm,width=7.5cm]{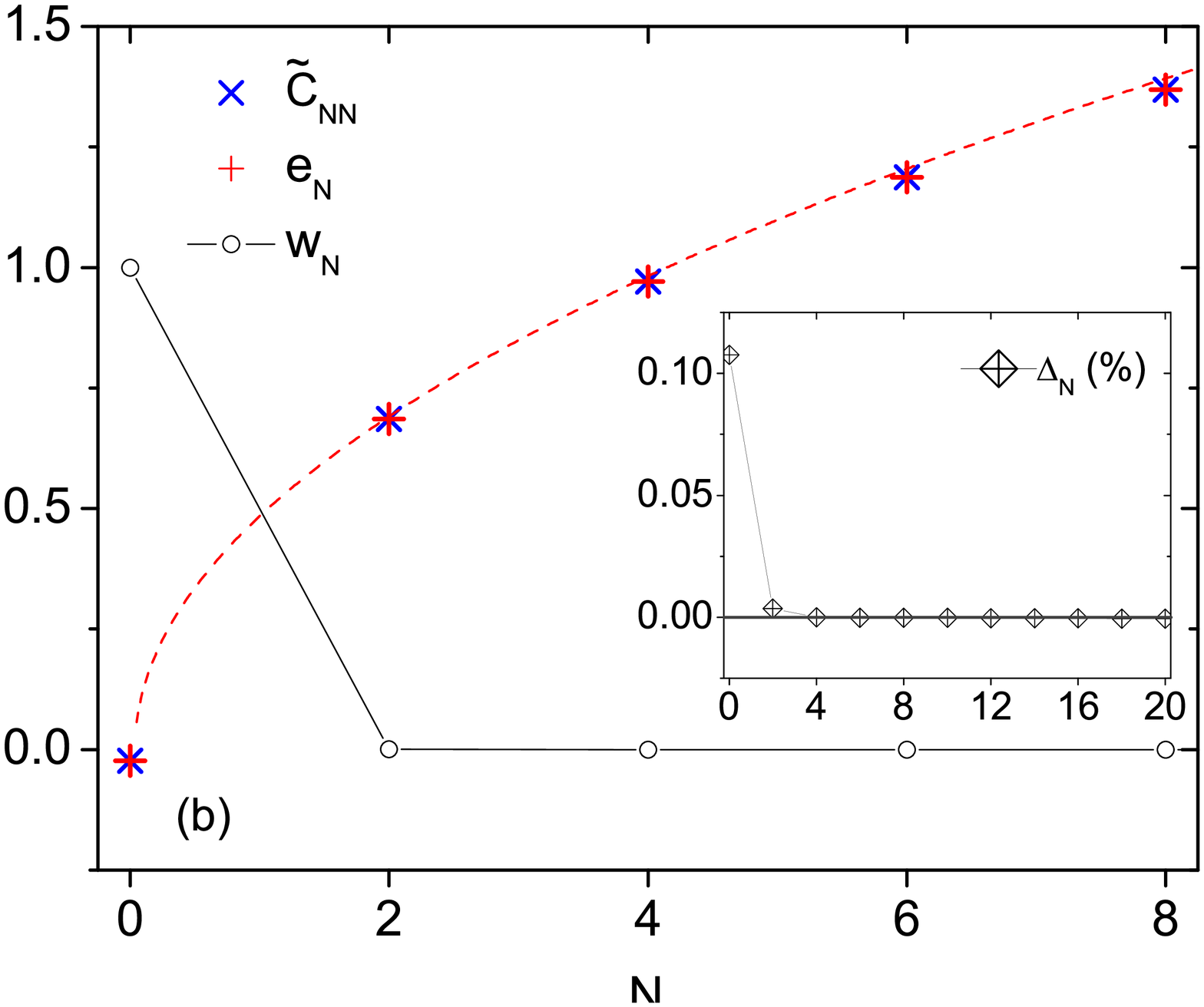}
\caption{(Color online) Same as Fig.1 except for
$^{40}$K(2D)-$^6$Li(3D) mixture. The dashed line in (a) denotes the
first resonance at $a_0/a_s=-0.02$.} \label{fig2}
\end{figure}

In addition, Fig.1 and Fig.2 show that the first resonance, which is
mainly due to the coupling between relative motion levels within the
lowest molecular channel($N=0$), always occurs at $a_s<0$ side
regardless of $u>1$ or $u<1$. The resonance position, however,
sensitively depends on the value of $u$, as shown by the vertical
dashed lines in Fig.1(a) and Fig.2(a). This phenomenon is closely
related to the distinct behaviors of $\tilde{C}_{NN}$ for different
$u$. On one hand, when $u\ll 1$ we can omit the $n_A$-dependence in
$\ln-$function in Eq.\ref{C_simp}, then using Eq.\ref{identity} we
obtain $\tilde{C}_{NN}\rightarrow0$ for any $N$. This implies that
when trapping the heavy atom, the resonance positions
[$(a_0/a_s)_{res}$] tend to highly aggregate around unshifted
position($a_0/a_s=0$). On the other hand, when $u\gg 1$,
$f_{N;n_A,k_z}$ is vanishingly small for finite $N$, then the first
term dominates in Eq.\ref{C_simp}.  This predicts resonances
approaching $a_s=0^-$. Only for large enough $N$ the rest terms in
Eq.\ref{C_simp} would dominate and predict resonances at $a_s>0$
side.

We note that the dependence of the first resonance position (at
$a_s<0$ side) on the mass ratio $u$ is in qualitative agreement with
that for 0D-3D mixtures\cite{Castin}. Actually we can gain the
physical insight of such feature from the analysis of interaction
potentials affected by the confinement. Suppose a square-well
interaction potential $U(r)$, which is $-V_0$ at $r<r_0$ and zero
otherwise, between A and B atoms. As $V_0$ increases, the first
scattering resonance occurs at the critical value
$V_c=(\pi/2)^2/(2\mu r_0^2)$ with $\mu$ the reduced mass. When A or
B is trapped and becomes localized, $\mu$ will be effectively
enhanced, which reduces the critical $V_c$ and gives new resonance
at $a_s<0$ side. $\mu$ and $V_c$ can be substantially modified if
the lighter atom is trapped, and the resonance position will move
far away to $a_s=0^-$ side. On the contrary, if the heavier atom is
trapped, $\mu$ and $V_c$ would be little affected by the trapping
potential, giving almost unshifted resonance near $a_s=\infty$. This
analysis leads to similar conclusions for the resonance scattering
in other mixed-dimensional systems, such as 1D-3D mixtures.

\section{Application to three-body problem}

Besides the two-body system, the general formulism of T-matrix
approach allows its straightforward extension to other few-body
systems. In this section we focus on a three-body system composed by
two-component fermions in a (rotating) harmonic trap. We shall first
present the formulism and then explore the interesting scattering
property and identify the ground state level crossing in this
system.

\subsection{Formulism}

We consider three fermions with one spin-$\downarrow$
($\mathbf{x}_1$) and two identical spin-$\uparrow$
($\mathbf{x}_2,\mathbf{x}_3$) in an isotropic harmonic trap.
According to Eqs.(\ref{A_matrix}-\ref{relative}), we transform the
vector
$\mathbf{X}=(\sqrt{2m_{\downarrow}}\mathbf{x}_1,\sqrt{2m_{\uparrow}}\mathbf{x}_2,\sqrt{2m_{\uparrow}}\mathbf{x}_3)$
to
$\mathbf{Y_{\pm}}=(\sqrt{2M_R}\mathbf{R},\sqrt{2\mu}\mathbf{r}_{\pm},\sqrt{2\mu}
\boldsymbol\rho_{\pm})$ by $\mathbf{Y_{\pm}}^T=A_{\pm}\mathbf{X}^T$.
Here $(\mathbf{R},\mathbf{r}_-,\boldsymbol\rho_-)$ and
$(\mathbf{R},\mathbf{r}_+,\boldsymbol\rho_+)$ are all Jacobi
coordinates, respectively corresponding to the effective mass
$M_R,\mu,\mu$; the CM coordinate $\mathbf{R}$ and its mass $M_R$
follow Eq.\ref{CM}; the other Jacobi coordinates are
\begin{eqnarray}
\mathbf{r_-}&=&\mathbf{x}_2-\mathbf{x}_1,\nonumber\\
\boldsymbol\rho_-&=&\frac{\sqrt{Mm_{\downarrow}}}{m_{\uparrow}+m_{\downarrow}}[\mathbf{x}_3-\frac{m_{\downarrow}\mathbf{x}_1+m_{\uparrow}\mathbf{x}_2}{m_{\uparrow}+m_{\downarrow}}],\label{r-rho}
\end{eqnarray}
with the same mass
$\mu=\frac{m_{\uparrow}m_{\downarrow}}{m_{\uparrow}+m_{\downarrow}}$;
the transfer matrix reads
\begin{equation}
A_-=\left(
    \begin{array}{ccc}
      \sqrt{\frac{m_{\downarrow}}{M}} & \sqrt{\frac{m_{\uparrow}}{M}} & \sqrt{\frac{m_{\uparrow}}{M}} \\
      -\sqrt{\frac{m_{\downarrow}}{m_{\uparrow}+m_{\downarrow}}} & \sqrt{\frac{m_{\uparrow}}{m_{\uparrow}+m_{\downarrow}}} & 0  \\
      -\frac{m_{\uparrow}}{\sqrt{M(m_{\uparrow}+m_{\downarrow})}} & -\sqrt{\frac{m_{\uparrow}m_{\downarrow}}{M(m_{\uparrow}+m_{\downarrow})}} & \sqrt{\frac{m_{\uparrow}+m_{\downarrow}}{M}}
    \end{array}
  \right);
\end{equation}
we further obtain $\boldsymbol\rho_+, \mathbf{r_+}$ by exchanging
$\mathbf{x}_2\leftrightarrow\mathbf{x}_3$ in
$\mathbf{r_-},\boldsymbol\rho_-$, and obtain $A_+$ by exchanging the
second and third column of $A_-$.

Taking advantage of the property of transfer matrix (Eq.\ref{AA}),
we can see that with the same trapping frequency $\omega$, all three
Jacobi coordinates
$(\mathbf{R},\mathbf{r}_{\pm},\boldsymbol\rho_{\pm})$ can be well
separated from each other. Independently one can also prove that the
total angular momentum is also separable as
$\sum_{i=1,2,3}\hat{L}_{\alpha}(\mathbf{x}_i)=\hat{L}_{\alpha}(\mathbf{R})+\hat{L}_{\alpha}(\boldsymbol\rho_{\pm})+\hat{L}_{\alpha}(\mathbf{r}_{\pm})\
(\alpha=x,y,z)$. Therefore for a trapped system with rotating
frequency $\Omega$ around z-direction, the relevant Hamiltonian in
the rotating frame reads
\begin{equation}
H(\boldsymbol\rho_{\pm},\mathbf{r_{\pm}})=H_0(\boldsymbol\rho_{\pm})+H_0(\mathbf{r_{\pm}})+U_0\delta(\mathbf{r_+})+U_0\delta(\mathbf{r_-}),
\end{equation}
here
\begin{equation}
H_0(\mathbf{r})=-\frac{\nabla^2_{\mathbf{r}}}{2\mu}+\frac{1}{2}\mu
\omega^2\mathbf{r}^2-\Omega L_z(\mathbf{r}).\label{H0_3body}
\end{equation}

The molecular state is defined with respect to Fermi statistics,
\begin{equation}
|\lambda\rangle=\frac{1}{\sqrt{2}}(|r_-=0, \lambda\rangle-|r_+=0,
\lambda\rangle). \label{lambda}
\end{equation}
Here the first and second $\lambda$ represent the identical energy
level $\{nlm\}$ for the motions of $\boldsymbol\rho_-$ and
$\boldsymbol\rho_+$ under Hamiltonian $H_0$. ($\{n\}$ and $\{lm\}$
are respectively the radial and azimuthal quantum number). Then we
obtain the $C-$matrix element as
\begin{equation}
C_{\mathbf{\lambda\lambda'}}=(\frac{1}{V}\sum_{\mathbf{k}}
\frac{1}{\epsilon_{\mathbf{k}}}+\sum_{\nu}
\frac{|\psi_{\nu}(0)|^2}{E-E_{\lambda}-E_{\nu}+i\delta})\delta_{\lambda\lambda'}-F_{\lambda\lambda'},\label{D_3body}
\end{equation}
with
\begin{eqnarray}
F_{\lambda\lambda'}&=& \langle r_-=0, \lambda|\hat{G}_0|r_+=0,
\lambda'\rangle\nonumber\\
&=&\int d\boldsymbol\rho\psi^*_{\lambda}(\boldsymbol\rho)
\psi_{\lambda'}(-\beta\boldsymbol\rho)\sum_{\nu}\frac{\psi_{\nu}(0)\psi_{\nu}(-\alpha\boldsymbol\rho)}{E-E_{\lambda'}-E_{\nu}+i\delta},\label{F}
\end{eqnarray}
here
\begin{equation}
\alpha=\frac{\sqrt{Mm_{\downarrow}}}{m_{\uparrow}+m_{\downarrow}},\
\ \ \
\beta=\frac{m_{\uparrow}}{m_{\uparrow}+m_{\downarrow}},\label{alpha-beta}
\end{equation}
and $\alpha^2+\beta^2=1$. To obtain Eq.\ref{F} we have inserted into
the Green function a complete set of eigen-states $\{\nu,\lambda'\}$
for the motions of $(\mathbf{r}_+,\boldsymbol\rho_+)$.
$F_{\lambda\lambda'}=F^*_{\lambda'\lambda}$ here induce the coupling
between different molecular levels, and non-zero
$F_{\lambda\lambda'}$ require azimuthal quantum number $\{lm\}$ be
conserved. Note that the off-diagonal coupling of molecular states
here is due to the many-body statistics, in contrary to the previous
two-body case which is due to the external trapping potentials. More
details regarding to the evaluation of Eq.\ref{D_3body} are
presented in Appendix \ref{3fermion}.

\subsection{Results}

In the first part of this section we use T-matrix method to analyze
the exotic scattering property of three fermions in different limits
of mass ratios and in different angular momenta channels. In the
second part, we present the energy spectrum and identify the energy
level crossing between different angular momenta states for the
(rotating) trapped system.

\subsubsection{Scattering property}

By analyzing Eqs.(\ref{D_3body},\ref{F}), we find nontrivial
scattering properties at two limits of mass ratio
$u=m_{\uparrow}/m_{\downarrow}$. Fig.\ref{fig_u} shows the schematic
plots of Jacobi coordinates $(\mathbf{r}_-,\boldsymbol\rho_-)$ in
both limits of $u\rightarrow0$ and $u\rightarrow\infty$.

\begin{figure}[ht]
\includegraphics[height=3.5cm,width=9cm]{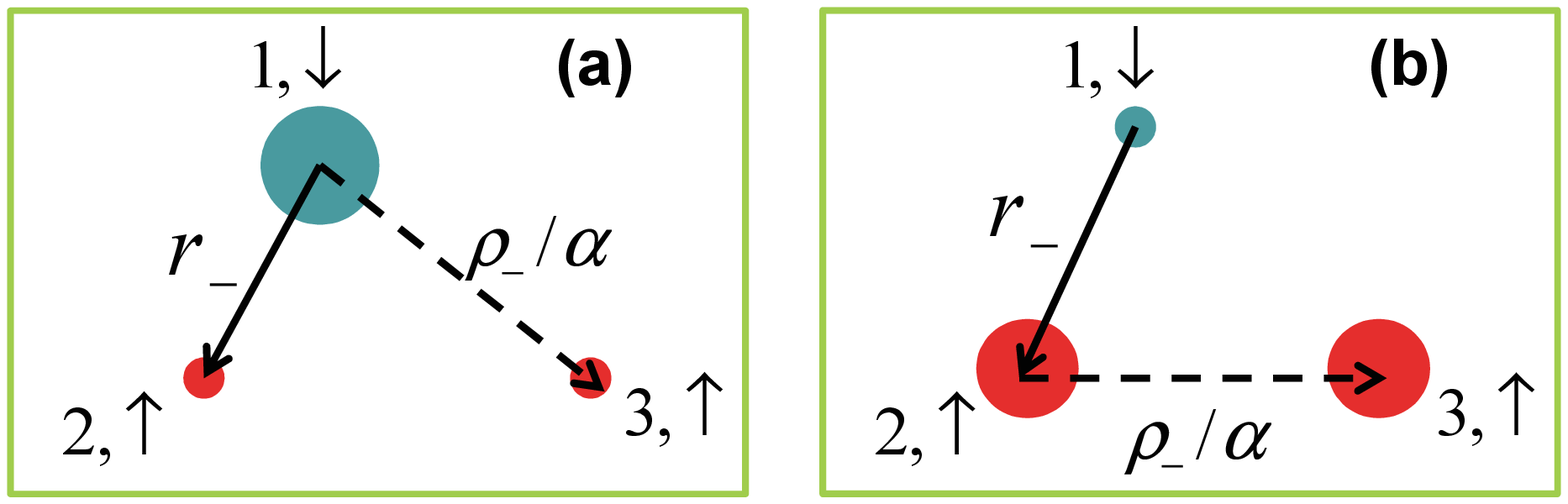}
\caption{(Color online) Jacobi coordinates
$(\mathbf{r}_-,\boldsymbol\rho_-)$ for three
fermions($\uparrow\uparrow\downarrow$) in the limit of
$u\rightarrow0$ (a) and $u\rightarrow\infty$ (b).
$u=m_{\uparrow}/m_{\downarrow}$ is the mass ratio. $\alpha$ is given
by Eq.\ref{alpha-beta}.} \label{fig_u}
\end{figure}

First, when $u\rightarrow0$ as shown by Fig.\ref{fig_u}(a),
$\alpha\rightarrow1,\ \beta\rightarrow0$, we have
\begin{equation}
F_{\lambda\lambda'}\sim\delta_{l,0}\delta_{l',0}.
\end{equation}
Therefore the diagonal C-matrix for $l\neq0$ indicates the
atom-dimer uncorrelated system with energy $E=E_{a}+E_{d}$.
Physically we can see from Fig.\ref{fig_u}(a) that, the dimer formed
by a light $\uparrow$ and heavy $\downarrow$ is almost equivalent to
single $\downarrow$ atom, so the other $\uparrow$ has s-wave
interaction with this dimer only when $l=0$. Here the heavy
$\downarrow$ dominates the whole physics.

Second, in the opposite limit when $u\rightarrow\infty$ as shown by
Fig.\ref{fig_u}(b), the result is completely different. We find in
this limit,
\begin{equation}
C_{\mathbf{\lambda\lambda'}}=(-\frac{1}{V}\sum_{\mathbf{k}}
\frac{1}{\epsilon_{\mathbf{k}}}-\sum_{\nu}
\frac{|\psi_{\nu}(0)|^2}{E-E_{\lambda}-E_{\nu}+i\delta}[1-(-1)^l])\delta_{\lambda\lambda'}.
\end{equation}
There are two direct consequences as follows.

(i)for odd $l$, $C_{\mathbf{\lambda\lambda'}}=\infty$, i.e.,
unphysical divergence in the high-energy space can not be properly
removed. This is exactly the evidence of Efimov effect for large $u$
where another short-range parameter is required to help fix the
three-body problem\cite{Efimov,Braaten06}.

(ii)for even $l$, $\hat{U}$ takes no effect and the system just
behaves like non-interacting. This result is consistent with that
obtained by Born-Oppenheimer approximation(BOA)\cite{NPA1979}. Under
BOA, the wavefunction is given by
\begin{equation}
\psi(\mathbf{x}_1,\mathbf{x}_2,\mathbf{x}_3)=[\varphi(|\mathbf{x}_2-\mathbf{x}_1|)+\gamma\varphi(|\mathbf{x}_3-\mathbf{x}_1|)]f(\mathbf{x}_2,\mathbf{x}_3),
\end{equation}
where the first part describes the light particle moving around two
static heavy particles, and $f(\mathbf{x}_2,\mathbf{x}_3)$ describes
for two heavy particles afterwards. By imposing Bethe-Peierls
boundary conditions one can find $\gamma=\pm 1$, and the energy of
the first part just depends on $|\mathbf{x}_2-\mathbf{x}_3|$.
Therefore the wavefunction is reduced to
\begin{equation}
\psi(\mathbf{x}_1,\mathbf{x}_2,\mathbf{x}_3)=[\varphi(\mathbf{|r_-|})\pm
\varphi(\mathbf{|r_+|})]f_1(\mathbf{x}_2-\mathbf{x}_3)f_2(\mathbf{R}),
\end{equation}
and then angular momentum $l$ is determined only by
$f_1(\mathbf{x}_2-\mathbf{x}_3)$. For $\gamma=1$, the Fermi
statistics require $l$ be odd; for $\gamma=-1$, $l$ is even but in
this case one can easily check that the resultant wavefunction
automatically get rid of the interaction, i.e., $\hat{U}\psi=0$.
Here Fermi statistics of two $\uparrow$ spins take the crucial role.

Note that the results presented in (i,ii) uniquely benefit from the
concept of renormalization and the procedure in momentum space to
eliminate the ultraviolet divergence. These analyses of scattering
properties for different mass ratios and different angular momenta
will be helpful to understand the ground state level crossing in the
following section.

\begin{figure}[ht]
\includegraphics[height=5cm,width=7.5cm]{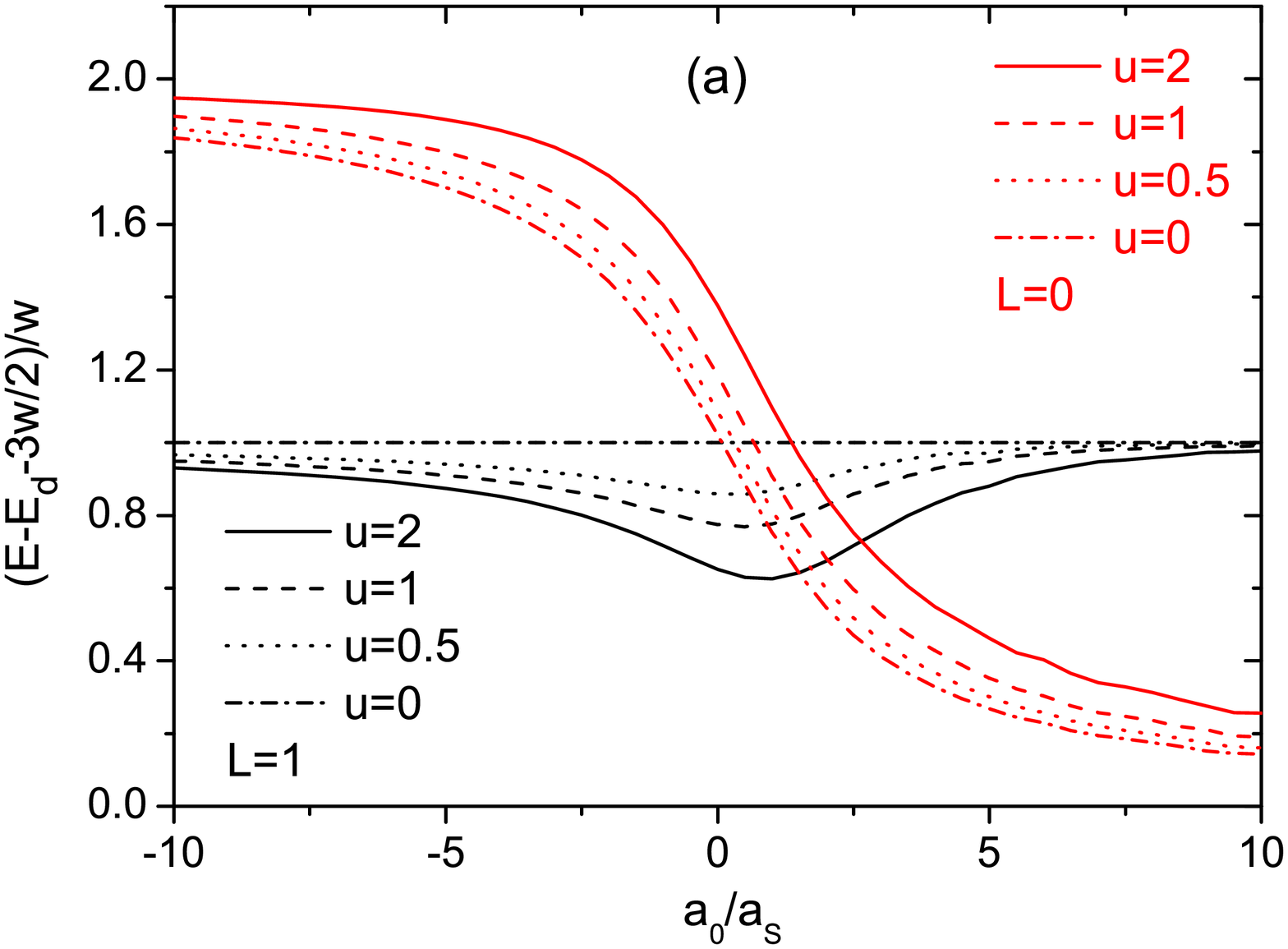}\\
\includegraphics[height=5cm,width=7.5cm]{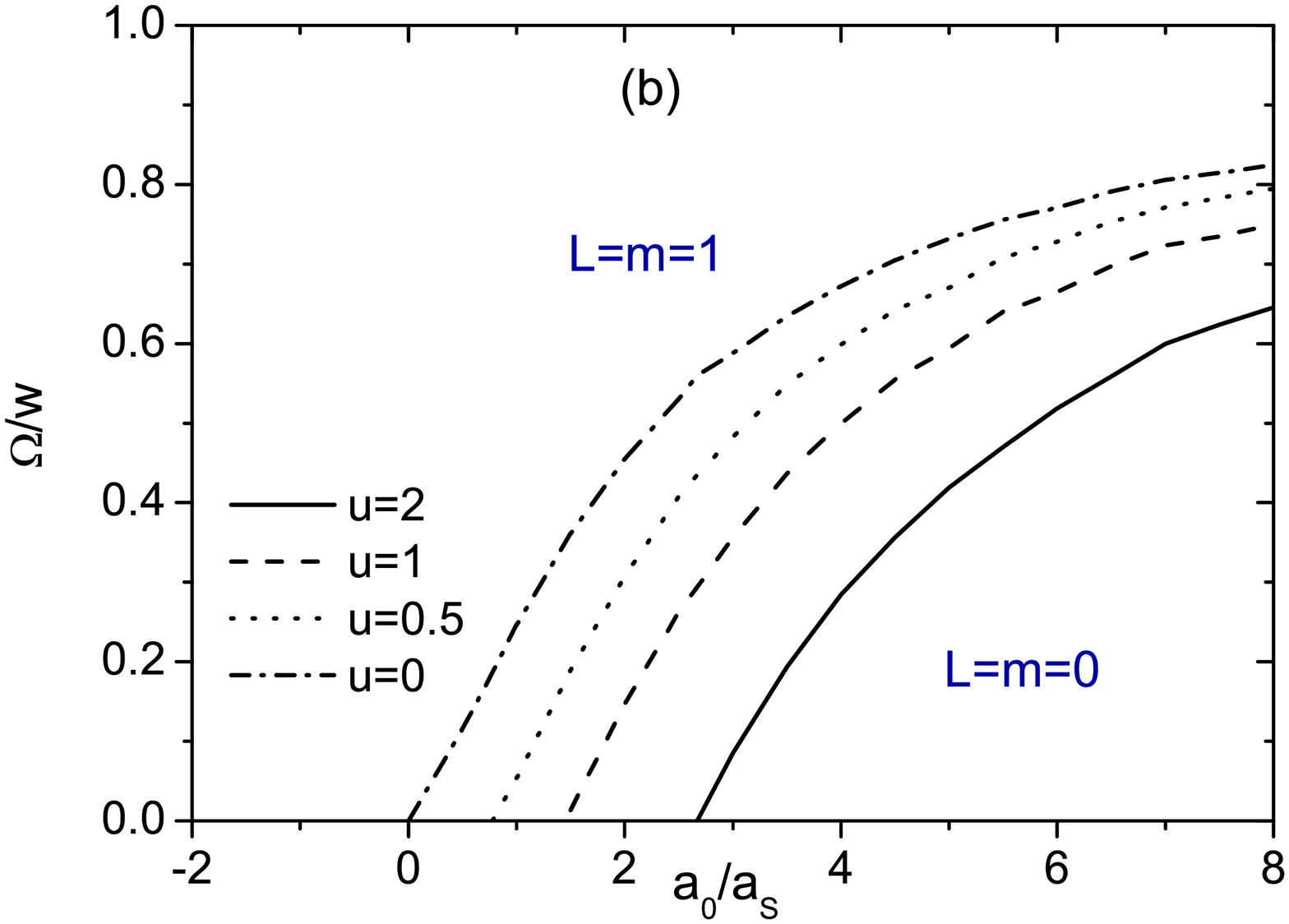}
\caption{(Color online) (a).Energy of three
fermions($\uparrow\uparrow\downarrow$) vs interaction strength in a
non-rotating isotropic harmonic trap, with $a_0=\sqrt{1/(\mu
\omega)}$ the confinement length. Different mass ratios
$u=m_{\uparrow}/m_{\downarrow}$ for total angular momentum
$l=0$(red) and $l=1$(black) are plotted. The energy is shifted by
the ground-state atom-dimer energy, $E_{d}+3\omega/2$($E_d$ is the
dimer energy). (b)Phase diagram in a rotating harmonic trap with
rotating frequency $\Omega$. The ground state is $|l=m=1\rangle$
above the curves and $|l=m=0\rangle$ below.}\label{QH}
\end{figure}

\subsubsection{Energy level crossing}

First, we identify the energy level crossing between different
angular momenta states for a non-rotating system. The energy
spectrum of a non-rotating system was previously studied for equal
mass\cite{Duan07, Liu10}, and unequal masses using Gaussian
expansion technique\cite{Stecher08} and adiabatic hyperspherical
method\cite{Rittenhouse10}. In Fig.\ref{QH}(a) we show the spectrum
for angular momenta $l=0,1$ and for different mass ratios using
T-matrix method. We also checked for higher $l\geq2$ and confirm
those states are less modified by the interaction and thus not shown
here.

The system in weak interacting limit($a_s\rightarrow0^-$) behaves as
non-interacting while in molecule limit($a_s\rightarrow0^+$) as a
single dimer plus an atom. This directly results in the inversion of
ground state from angular momentum $l=1$ to $l=0$ as $1/a_s$
increases. As shown in Fig.\ref{QH}(a), the inversion is denoted by
the energy level crossing, and the position of level crossing
closely depends on mass ratio $u$. As expected, when $u$ increases
from $0$ to $\infty$, all energy levels with even-$l$ move upwards
and the system evolves from decoupled atom-dimer(except for $l=0$)
to three atoms that are immune from interactions; while all odd-$l$
move downwards until Efimov physics show up and invalidate the
present T-matrix method. Therefore by increasing $u$, the position
of level crossing will move to strong coupling side as shown in
Fig.\ref{QH}(a). Intuitively, one can also attribute this to the
enhanced s-wave repulsion between atom and dimer\cite{Petrov1}.

In unitary limit, our numerical results are in good accordance with
those obtained by using hyperspherical coordinates. Previously,
hyperspherical coordinate method has been applied to a trapped
system with equal mass\cite{Werner061,Werner062}. In Appendix
\ref{appendix_c} we extend this method to arbitrary mass ratios.
Note that the maximum mass ratio considered in Fig.3 is much less
than critical value, $u_c=13.6$\cite{Petrov1}, for the emergence of
Efimov state in $l=1$ channel (as also predicted by Eq.\ref{l1} when
setting $s=0$). In this regime, as the matrix size increases we get
convergent result for the energy spectrum.

Finally, we present the ground state for the trapped system with
rotation($\Omega>0$). By comparing energies of all different angular
momenta $l$ we obtain the ground state as shown in Fig.\ref{QH}(b).
We find $l=m=1$ state is gradually favored by the rotation. The
energy gain of this state is analyzed to be partly from the
reduction of kinetic energy with respect to $l=0$, and partly from
the avoided s-wave repulsion between atom and dimer. As $\Omega$
increases, the system evolves to the atom-dimer quantum Hall state;
particularly at $\Omega=\omega$, all states with odd-$l$ degenerate.
Finally we expect above quantum Hall physics of fermionic system
could be studied in experiment, as recently realized in a rotating
few-body bosonic system\cite{Gemelke10}.

\section{Summary}

In conclusion, we present a systematic T-matrix approach to solve
few-body problems with contact interactions in the field of
ultracold atoms. Taking advantage of zero-range interactions, the
key ingredient of the present T-matrix method is to project the
problem to a subspace that is expanded by orthogonal molecular
states, and meanwhile take careful considerations of the
renormalization for relative motions. This method successfully
unifies the calculations of various physical quantities in a single
framework, including the bound state solutions, effective scattering
lengths and reduced interactions in the lower dimension.

We present two applications of this approach, namely, two-body
scattering resonances in 2D-3D mixed dimensions and properties of
three fermions($\uparrow\uparrow\downarrow$) in a 3D (rotating)
trap. For the two-body problem, we show that T-matrix provides a
physically transparent way to understand the mechanism of induced
scattering resonances. Besides, it also gives explicit expressions
for the resonance positions and widths. Due to the separate
treatment of relative motions from CM motions, each resonance can be
determined accurately by only considering a few related matrix
elements. For the three-body problem, T-matrix enables us to
identify exotic scattering properties of three fermions in different
angular momentum channels and with different mass ratios.  In a
rotating system, these properties provide important hints for the
quantum Hall transition from zero to finite angular momentum state.
Overall, the external confinements, mass ratios, and
bosonic/fermionic statistics all play important roles and give rise
to very rich phenomenon in these few-body systems.

The author thanks Hui Zhai, Shina Tan, Fei Zhou, Hui Hu and Doerte
Blume for useful discussions, and Jason Ho for valuable suggestions
on the manuscript. This work is supported by Tsinghua University
Basic Research Young Scholars Program and Initiative Scientific
Research Program and NSFC under Grant No. 11104158. The author would
like to thank the hospitality of the Institute for Nuclear Theory at
University of Washington, where this work is finally completed
during the workshop on "Fermions from Cold Atoms to Neutron Stars"
in the spring of 2011.

\bigskip

\appendix

\section{Construction of individual molecular
state}\label{indi mole}

In this appendix, we show how to construct an individual molecular
state in the most efficient way.
Let us consider a system of $N$ particles with masses $m_1,...m_N$
and coordinates $\mathbf{x}_{1},...\mathbf{x}_{N}$. For a general
case, the molecular state is written as
\begin{equation}
|\mathbf{x}_1=\mathbf{x}_2,\lambda=\{N,n_3,n_4,...n_N\}\rangle.
\label{general}
\end{equation}
In coordinate space it can be factorized as
$\Phi_{N}(\mathbf{X})\prod_{i=3}^N\phi_{n_i}(\mathbf{x}_i)$, where
$\phi_{n_i}(\mathbf{x}_i)$ is the eigen-state of single-particle
Hamiltonian
\begin{equation}
\hat{H}_0(\mathbf{x}_i)=-\frac{\nabla_{\mathbf{x}_i}^2}{2m_i}+V_{T,i}(\mathbf{x}_i)
\end{equation}
and $\Phi_{N}(\mathbf{X})$ the eigen-state of
\begin{equation}
\hat{H}_0(\mathbf{X})=-\frac{\nabla_{\mathbf{X}}^2}{2(m_1+m_2)}+V_{T,2}(\mathbf{X})+V_{T,2}(\mathbf{X}).
\end{equation}
Here $V_{T,i}$ is the trapping potential for the $i-$th particle.
The overlap between the molecular state(Eq.\ref{general}) and
N-particle state($\prod^N_{j=1}|l_j\rangle$) is
\begin{equation}
\langle
\mathbf{x}_1=\mathbf{x}_2,\lambda|l_1,l_2,l_3,...l_N\rangle=\gamma_{N;l_1,l_2}\prod_{i=3}^N\delta_{n_il_i}
\end{equation}
with
\begin{equation}
\gamma_{N;l_1,l_2}=\int
d\mathbf{X}\Phi_N^*(\mathbf{X})\phi_{l_1}(\mathbf{X})\phi_{l_2}(\mathbf{X}).
\end{equation}
The Green function term in Eq.\ref{f_matrix2} can then be computed
efficiently as
\begin{eqnarray}
&&\langle r_I=0,\lambda|\hat{G}_0(E)|r_{I'}=0,\lambda'
\rangle\nonumber\\
&=&\sum_{l_1...l_N}\frac{\langle
r_I=0,\lambda|l_1...l_N\rangle\langle
l_1...l_N|r_{I'}=0,\lambda'\rangle}{E-(E_{l_1}+...+E_{l_N})+i\delta}.\label{D_unsepa2}
\end{eqnarray}

To this end we have shown a general way to construct an individual
molecular state. Furthermore, for special trapping potentials which
enable the decoupling of CM motion from other motions, it is
convenient to remove the CM motion from the problem and transform
the effective coordinate vector
\begin{equation}
\mathbf{X}=(\sqrt{2m_1}\mathbf{x}_1,\sqrt{2m_2}\mathbf{x}_2,\sqrt{2m_3}\mathbf{x}_3,...\sqrt{2m_{N}}\mathbf{x}_N)
\label{X}
\end{equation}
to the Jacobi coordinates
\begin{equation}
\mathbf{Y}=(\sqrt{2M_R}\mathbf{R},\sqrt{2\mu}\mathbf{r},
\sqrt{2\nu_1}\boldsymbol\rho_1,...\sqrt{2\nu_{N-2}}\boldsymbol\rho_{N-2})\label{Y}
\end{equation}
by a matrix equation
\begin{equation}
\mathbf{Y}^T=A\mathbf{X}^T,
\end{equation}
with A-matrix element
\begin{eqnarray}
A_{ij}=\left\{\begin{array}{l} \sqrt{m_j/M_N},\ \  (i=1) \\
\sqrt{M_{i-1}/M_i},\ \ (i=j>1)\\ -\sqrt{m_im_j/(M_{i-1}M_i)},\ \
 (i>j\geq1) \\ 0, \ \ \textrm{all else}
\end{array}\right. \label{A_matrix}
\end{eqnarray}
here $M_j=\sum_{i=1}^j m_i$.

For the CM motion, we have the coordinate and the mass
\begin{equation}
\mathbf{R}=\sum_{i=1}^N \frac{m_i\mathbf{x}_i}{M_N}, \ \ \
M_R=\sum_{i=1}^N m_i;\label{CM}
\end{equation}
for other motions, we take the unique choice as
\begin{eqnarray}
\sqrt{2\mu}\mathbf{r}&=&\sqrt{\frac{2m_1m_2}{m_1+m_2}}(\mathbf{x}_2-\mathbf{x}_1),\\
\sqrt{2\nu_j}\boldsymbol\rho_j&=&\sqrt{\frac{2m_{i+2}M_{i+1}}{M_{i+2}}}(\mathbf{x}_{j+2}-\sum_{i=1}^{j+1}
\frac{m_i\mathbf{x}_i}{M_{j+1}}).\label{relative}
\end{eqnarray}
According to Eq.\ref{A_matrix}, A-matrix satisfies
\begin{equation}
AA^T=A^TA=I,\label{AA}
\end{equation}
here $I$ is identity matrix; this gives
\begin{equation}
d\mathbf{R}d\mathbf{r}\prod^{N-2}_j d\rho_j=\prod^N_i d\mathbf{x}_i
\end{equation}
and
\begin{equation}
\nabla_{\mathbf{X}}\nabla^T_{\mathbf{X}}=\nabla_{\mathbf{Y}}\nabla^T_{\mathbf{Y}},\
\ \mathbf{X}\mathbf{X}^T=\mathbf{Y}\mathbf{Y}^T. \label{separate}
\end{equation}
Therefore $M_R, \mu, \nu_1,...\nu_{N-2}$ can be considered as the
effective mass respectively for the motion of $\mathbf{R},
\mathbf{r}, \boldsymbol\rho_1,...\boldsymbol\rho_{N-2}$. By
decomposing the trapping potential to be
\begin{equation}
\sum_iV_{T,i}(\mathbf{x}_i)=V_T(\mathbf{R})+V_T(\mathbf{r},\boldsymbol\rho_1,...\boldsymbol\rho_{N-2}),
\end{equation}
we choose the molecular state $|r=0,\lambda\rangle$ such that its
real-space wavefunction
$\phi_{\lambda}(\boldsymbol\rho_1,...\boldsymbol\rho_{N-2})$ is the
eigen-state of the following Hamiltonian for $N-2$ particles
\begin{equation}
\hat{H}^{N-2}_0=-\sum_{j=1}^{N-2}\frac{\nabla_{\boldsymbol\rho_j}^2}{2\nu_j}+V_T(r=0,\boldsymbol\rho_1,...\boldsymbol\rho_{N-2}).
\end{equation}
The Green function in Eq.\ref{f_matrix2} is obtained by inserting
eigen-states of the following Hamiltonian for $N-1$ particles
\begin{equation}
\hat{H}^{N-1}_0=-\frac{\nabla_{\mathbf{r}}^2}{2\mu}-\sum_{j=1}^{N-2}\frac{\nabla_{\boldsymbol\rho_j}^2}{2\nu_j}+V_T(\mathbf{r},\boldsymbol\rho_1,...\boldsymbol\rho_{N-2}).
\end{equation}
Compared with Eq.(\ref{general}-\ref{D_unsepa2}) for a general case,
the degree of freedom here is further reduced by one particle.

\section{Derivation of Eq.\ref{f_matrix}}\label{C_derive}

To facilitate the derivation of Eq.\ref{f_matrix}, we assume there
is only one pair of particles interacting with
$U_I\delta^3(\mathbf{r}_I)$, and we have
\begin{equation}
\langle
\mathbf{r}_{I},\lambda|\hat{U}=U_I\delta^3(\mathbf{r}_I)\langle
r_{I}=0,\lambda|.
\end{equation}
Using the Lippmann-Schwinger equation or equivalently
\begin{equation}
T=U+UG_0T,
\end{equation}
and together with Eq.\ref{f} we obtain
\begin{eqnarray}
\sum_{I'\lambda'}f^{I'}_{\lambda'}|r_{I'}=0,\lambda'\rangle=U|\psi_0\rangle+UG_0\sum_{I'\lambda'}f^{I'}_{\lambda'}|r_{I'}=0,\lambda'\rangle.
\label{f_matrix3}
\end{eqnarray}
The inner product with $\langle r_{I}=0,\lambda_I|$ gives
\begin{eqnarray}
&&\sum_{I'\lambda'}f^{I'}_{\lambda'}(\frac{1}{U_I}\delta_{II'}\delta_{\lambda\lambda'}-\langle
r_I=0,\lambda|\hat{G}_0|r_{I'}=0,\lambda' \rangle)\nonumber\\
&&\ \ \ \ \ \ \ \ =\langle r_I=0,\lambda |\psi_0 \rangle.
\label{f_matrix4}
\end{eqnarray}
Here we have extracted the most singular terms as characterized by
$\delta^3(\mathbf{r}_I)$. After further renormalizing the bare
interaction $U_I$ we obtain Eq.\ref{f_matrix}.

Above derivation can be generalized to the case when the molecular
state $|r_I=0,\lambda_I \rangle$ is the superposition of many
individual ones according to bosonic or fermionic statistics. After
a proper combination of the resulted individual equations, one can
equally obtain Eq.\ref{f_matrix}.

\section{Effective scattering in 2D-3D mixture}\label{a_eff}

In this appendix we study two-body scattering in 2D-3D mixed
dimensions(see also Section IIIB).

First, we relate the effective scattering length to T-matrix element
between zero-energy scattering states. From Eq.\ref{psi}, the
two-body wavefunction in coordinate space reads
\begin{eqnarray}
&&\psi(\boldsymbol\rho_{AB},z_A,z_B)=\psi_0(\boldsymbol\rho_{AB},z_A,z_B)+\sum_{n_A}\sum_{
\mathbf{k}_{\rho},k_z} \phi_{n_A}(z_A,a_0)\nonumber\\
&&
\frac{1}{\sqrt{V}}\frac{e^{i\mathbf{k}_{\rho}\cdot\boldsymbol\rho_{AB}}e^{ik_z
z_{B}}}{E-n_A\omega_A-\frac{\mathbf{k}_{\rho}^2}{2\mu}-\frac{k_{z}^2}{2m_B}+i\delta}
\langle \mathbf{k}_{\rho}, n_A, k_z|T|\psi_0 \rangle,
\label{psi_appendix}
\end{eqnarray}
with $\boldsymbol\rho_{AB}$ the relative coordinate of A and B in xy
plane and $z_A,z_B$ their respective coordinate in z-direction; $E$
is counted from the zero-point energy $\omega_A/2$. For large
separations, both $e^{i\mathbf{k}_{\rho}\cdot\boldsymbol\rho_{AB}}$
and $e^{ik_z z_{B}}$ oscillate far more rapidly in k-space than
T-matrix term. Thus in the limits of $E\rightarrow 0^+$ and
$|\boldsymbol\rho_{AB}|,\ |z_B|\rightarrow\infty$, we can specify
$\mathbf{k}_{\rho}=0,\ k_z=0$ in all T-matrix elements and reduce
Eq.\ref{psi_appendix} to
\begin{eqnarray}
&&\psi(\boldsymbol\rho_{AB},z_A,z_B)\rightarrow
\phi_0(z_A)(1-\frac{\mu}{2\pi
d_{AB}}V\langle\psi_0|T|\psi_0\rangle)\nonumber\\
&&\ \ -\sum_{n_A>0}\phi_{n_A}(z_A,a_0) \frac{\mu}{2\pi
d_{AB}}e^{-\kappa_{n_A}d_{AB}}V\langle0,n_A,0|T|\psi_0\rangle\nonumber\\
&&\rightarrow \phi_0(z_A,a_0)(1-\frac{a_{\rm eff}}{d_{AB}}), \ \ \
(d_{AB}\rightarrow\infty)  \label{psi_2d3d}
\end{eqnarray}
with $\kappa_{n_A}=\sqrt{2m_Bn_A\omega_A}$,
$d_{AB}=\sqrt{\frac{\mu}{m_B}\boldsymbol\rho_{AB}^2+z_B^2}$, and
$a_{\rm eff}$ directly related to T-matrix element as given by
Eq.\ref{aeff-T}.

In Eq.\ref{aeff-as}, $e_{\nu}$ and $W_N$ can be obtained from the
diagonalization of the following matrix
\begin{eqnarray}
\tilde{C}_{NN'}&=&\frac{2\pi a_0}{\mu} \Big\{
\frac{1}{V}\sum_{\mathbf{k}}
\frac{1}{\mathbf{k}^2/(2\mu)}\delta_{NN'} + \nonumber\\
&&\sum
_{n_A}\sum_{k_z,\mathbf{k}_{\rho}}\frac{1}{V}\frac{f_{N;n_A,k_z}f^*_{N';n_A,k_z}}{E-n_A\omega_A-\frac{\mathbf{k}_{\rho}^2}{2\mu}-\frac{k_{z}^2}{2m_B}+i\delta}
\Big\}, \label{C_matrix}
\end{eqnarray}
with
\begin{equation}
f_{N;n_A,k_z}=\int_{-\infty}^{+\infty}dZ
\phi^*_N(Z,\overline{a})\phi_{n_A}(Z,a_0)e^{ik_zZ}.\label{f-function}
\end{equation}
Using the exact identity
\begin{equation}
\sum_{n_A}|f_{N;n_A,k_z}|^2=1, \label{identity}
\end{equation}
we eliminate the logarithmic divergence when summing over
$\mathbf{k}_{\rho}$ and finally simplify Eq.\ref{C_matrix} to be
\begin{eqnarray}
\tilde{C}_{NN'}&=&\frac{1}{\pi}\int_0^{\infty}d\tilde{k}_z \Big\{
-\ln(\tilde{k}_z^2)\delta_{NN'} + \sum
_{n_A}\nonumber\\
&&f_{N;n_A,k_z}f^*_{N';n_A,k_z}\ln(\frac{\tilde{k}_z^2}{1+u}+n_A\frac{2u}{1+u})
\Big\}. \label{C_simp}
\end{eqnarray}

In practical calculations, we set the cutoff of $n_A$ to be $1000$
for K-Rb($m_A<m_B$) and $700$ for K-Li($m_A>m_B$) mixture, depending
on the convergence in terms of $n_A$ in each case. The cutoff of
$\tilde{k}_z(=k_za_0)$ is chosen within $20\sim30$ to check the
insensitive dependence on the cutoff and ensure the accuracy of the
integration.

\section{Evaluation of Eq.\ref{D_3body}} \label{3fermion}

The eigen-state of Hamiltonian (Eq.\ref{H0_3body}) is
\begin{equation}
\psi_{nlm}(r,\theta,\phi)=\mathcal{N}_{nl}a_0^{-3/2}(\frac{r}{a_0})^l
e^{-r^2/(2a_0^2)}L_n^{l+\frac{1}{2}}(\frac{r^2}{a_0^2})Y_{lm}(\theta,\phi)
\end{equation}
where $\mathcal{N}_{nl}=\sqrt{\frac{2n!}{\Gamma(n+l+3/2)}}$,
$a_0=\sqrt{1/(\mu w)}$; $L_n^{l+\frac{1}{2}}$ is the generalized
Laguerre polynomial; the corresponding eigen-energy is
$E_{nlm}=(2n+l+\frac{3}{2})\omega-m\Omega$.

The eigen-solution($E$) of the interacting system is determined by
Eq.\ref{Eb}, or equivalently (for given $\{lm\}$)
\begin{equation}
Det(\frac{a_0}{a_s}\delta_{nn'}-\tilde{C}_{nn'})=0,\label{det_3body}
\end{equation}
where $\tilde{C}_{nn'}=A_{n}\delta_{nn'}-\tilde{F}_{nn'}$, with
\begin{equation}
A_{n}=\lim_{\Lambda\rightarrow\infty} \Big[
\frac{4\sqrt{\Lambda}}{\pi}+\sum_{n''=0}^{\Lambda}
\frac{\frac{2}{\sqrt{\pi}}\frac{(2n''+1)!!}{(2n'')!!}}{\tilde{E}-\tilde{E}_{nlm}-\tilde{E}_{n''00}}\Big],
\end{equation}
\begin{equation}
\tilde{F}_{nn'}=\sum_{n''}\frac{B_{nn'n''}}{\tilde{E}-\tilde{E}_{nlm}-\tilde{E}_{n''00}},
\end{equation}
\begin{eqnarray}
B_{nn'n''}&=&(-\beta)^{l}\mathcal{N}_{nl}\mathcal{N}_{n'l}\frac{2}{\sqrt{\pi}}\int_0^{\infty}dx
x^{2l+2}e^{-x^2}\nonumber\\
&&L_{n}^{l+\frac{1}{2}}(x^2)L_{n'}^{l+\frac{1}{2}}(\beta^2x^2)L_{n''}^{\frac{1}{2}}(\alpha^2x^2).
\end{eqnarray}
Here $\tilde{E}=E/\omega$. In our numerical simulations, the typical
size of the matrix for diagonalization is $50\times50$. When sum
over intermediate states, the actual cutoff of $n''$, namely $n^c$,
depends on the values of $\{n,n'\}$, which is constrained by
$B_{nn'n^c}\leq 10^{-4}$ in practical calculations.

\section{Unitary three-fermions with arbitrary mass ratios}\label{appendix_c}

In unitary limit, it is convenient to use hyperspherical coordinates
to express the three-body wavefunction\cite{Braaten06, Werner061,
Werner062} as
\begin{equation}
\psi(\mathbf{r}_{\pm},\boldsymbol\rho_{\pm})=\frac{F(R)}{R^2}(1-P_{23})\frac{\varphi(\xi)}{\sin(2\xi)}Y_{lm}(\hat{\rho}_-),
\label{hyper_psi}
\end{equation}
with $\mathbf{r}_{\pm},\ \ \boldsymbol\rho_{\pm}$ given by
Eq.\ref{r-rho}; $P_{23}$ is the permutation operator for identical
fermions $2$ and $3$;
\begin{equation}
R=\sqrt{\frac{\mathbf{r}^2_{\pm}+\boldsymbol\rho^2_{\pm}}{2}},\ \ \
\ \xi=\arctan\frac{|\mathbf{r}_-|}{|\boldsymbol\rho_-|},
\end{equation}
are respectively the hyperradius and hyperangle. Using this ansatz,
we obtain two decoupled Schrodinger equations for $R$ and $\xi$,
\begin{equation}
\big[-\frac{\hbar^2}{4\mu}(\frac{d^2}{dR^2}+\frac{1}{R}\frac{d}{dR})+\frac{\hbar^2s^2}{4\mu
R^2}+\mu \omega^2R^2\big]F(R)=EF(R); \label{f_R}
\end{equation}
\begin{equation}
\big[-\frac{d^2}{d\xi^2}+\frac{l(l+1)}{\cos^2\xi}\big]\varphi(\xi)=s^2\varphi(\xi).
\label{f_phi}
\end{equation}
On the other hand, the Bethe-Peierls boundary condition applying to
Eq.\ref{hyper_psi} gives
\begin{equation}
\varphi'(0)-\frac{(-1)^l}{\alpha\beta}\varphi(\arccos\beta)=0,
\label{hyper_BP}
\end{equation}
with $\alpha,\ \beta$ given by Eq.\ref{alpha-beta}. Combined with
Eq.\ref{f_phi}, Eq.\ref{hyper_BP} generates the following equations
for $l=0$ and $l=1$ (assuming all $s$ are real and positive),
\begin{equation}
s\cos(\frac{\pi}{2}s)+\frac{1}{\alpha\beta}\sin(s\arcsin\beta)=0,\ \
\ (l=0) \label{l0}
\end{equation}

\begin{eqnarray}
(1-s^2)\sin(\frac{\pi}{2}s)&=&\frac{1}{\alpha\beta}[s\cos(s\arcsin\beta)-\frac{\alpha}{\beta}\sin(s\arcsin\beta)]\nonumber\\
&&  \ \ \ \ \ \ \ \ \ \ \ \ \ \ \ \ \ \ \ \ (l=1).\label{l1}
\end{eqnarray}
which are the most relevant equations to the phase transition
discussed in Section IVB. For equal mass $\mu=m/2$, above equations
reduce to those in Ref.\cite{Werner061,Werner062}.

Note that the validity of above assumption ($s>0$) in deriving
Eqs.(\ref{l0},\ref{l1}) closely depends on the mass ratio $u$ and
the angular momentum $l$. For any given $l$, the hyperangular
equation (\ref{f_phi}) and condition (\ref{hyper_BP}}) provide the
solutions for $s$, which has double degenaracy($s$ and $-s$). For
$l=1$ channel, the zero-range model would provide universal energy
solutions for $u<8.62$ (with $s>1$)\cite{Nishida}. When
$8.62<u<13.6\ (0<s<1)$, the energy solution will depend on the
details of interacting potentials and 3-body resonance might happen
by tuning the potentials\cite{Blume_prl,Blume_pra}. When $u>13.6\
(s^2<0)$ the imaginary $s$ indicates the Efimov physics with
infinite number of shallow trimers\cite{Petrov1}.

In the actual computation of energy spectrum using T-matrix, for
nearly all mass ratios below $13.6$ one would get the energy
spectrum with good convergence, as the matrix size of
Eq.\ref{det_3body} is increased. The signal of 3-body
resonance(occurs around $u=12.3$ with $s=1/2$) is vanishing weak due
to the infinitesimal width produced by zero-range
model\cite{Blume2}. Therefore in the regime of $u<13.6$ one can just
consider the $s>0$ solution. Then the hyperradius equation
(\ref{f_R}) gives the energy $E=(2n+s+1)\omega$ with $n(=0,1...)$ a
semi-positive integer. In the rotating frame, the energy is further
shifted by $-m\Omega$, with $m$ the magnetic quantum number and
$\Omega$ the rotating frequency.


\begin{thebibliography}{}
\bibitem{note}$g_{\rm eff}$ here is the modified "bare"
interaction for open channel particles, which in this article is
essentially referred to the reduced 1D coupling constant. In 2D,
further regulariztion is needed to get physical one due to the
logrithmic divergence in the renormalization procedure.
\bibitem{insert}The standard way is to insert a complete set of
states as $\sum_{I\lambda}|r_I,\lambda\rangle\langle r_I,\lambda|$.
The zero-range property of interaction simplifies the situation as
inserting a set of molecular states (see also proof in Appendix
\ref{C_derive}).
\bibitem{convergence}The divergence of each term in Eq.\ref{CE} is resulted from the integration at high-energy regime, where the integral behaves
asymptotically as $\rho(E_n)|\phi_n(r=0)|^2/E$, with $\rho(E)$ the
density of state and $\phi_n(r)$ the wavefunction of relative
motion. In the extremely high-energy space, $W$ is essentially
determined by kinetic terms and therefore is identical to that in
free space. This in turn ensures the exact cancellation of the two
divergences in Eq.\ref{CE}.
\bibitem{aeff-T_proof}See Eq.\ref{aeff-T} for 2D-3D mixture and its proof in Appendix \ref{a_eff}. The statement
holds
true for effective scattering in the generalized three spatial
dimensions, such as in 0D-3D, 1D-3D and 1D($//$)-2D($\perp$)
mixtures.
\end{thebibliography}
\end{document}